\documentclass[11pt, a4paper]{article}

\usepackage{amsmath}
\usepackage{amssymb}
\usepackage{amsthm}
\usepackage[numbers]{natbib}
\usepackage{graphicx}
\usepackage{url}
\usepackage{xcolor}
\usepackage{times}
\usepackage{bm}
\usepackage{booktabs}
\usepackage{array}
\newcolumntype{P}[1]{>{\raggedright\arraybackslash}p{#1}}

\newcommand{\rr}{\mathbb{R}}

\DeclareMathOperator{\sym}{sym}
\DeclareMathOperator{\blkdiag}{blkdiag}
\DeclareMathOperator*{\argmin}{arg\,min}
\newcommand{\T}{^\top}

\newtheorem{proposition}{Proposition}

\begin{document}

\title{Learning monotone game-theoretic linear MPC}

\author{%
Alberto Bemporad\\
IMT School for Advanced Studies Lucca, Italy\\
\texttt{alberto.bemporad@imtlucca.it}
\and
Tatiana Tatarenko\\
Technische Universit\"at Darmstadt, Germany\\
\texttt{tatiana.tatarenko@tu-darmstadt.de}
}

\maketitle

\begin{abstract}%
We propose a method to learn a game-theoretic model predictive control (GT-MPC) law by imitation from input and output data of a running multi-agent system. By interpreting the recorded inputs as the first moves of the equilibria of a game played by the agents,
we learn the GT-MPC weights under hard input constraints and soft output constraints,
without requiring the agents to reveal their objectives. We also impose that the resulting generalized Nash equilibrium (GNE) problem is strongly monotone, which ensures the existence and uniqueness of the equilibrium and, hence, that the learned GT-MPC law is well defined and can be evaluated efficiently online by variational GNE-seeking algorithms for linear-quadratic games. Moreover, strong monotonicity implies that the equilibrium is almost everywhere differentiable with respect to the references and the weights, which enables automatic differentiation for gradient-based optimization. In addition, we promote closed-loop stability of the unconstrained equilibrium feedback by incorporating a penalty that involves a jointly learned quadratic Lyapunov function. Learning a GT-MPC formulation enables us to explain the agents' objectives, reproduce their behavior in simulation
in new scenarios, optimize their set-points, design mechanisms, and enforce additional constraints.
A numerical example illustrates the approach.
\end{abstract}


\section{Introduction}
\label{sec:intro}

In game-theoretic model predictive control (GT-MPC), several agents act on a common dynamical system, each optimizing a personal objective that depends on the other agents' moves under local and shared constraints, and applying the first move of the resulting generalized Nash equilibrium (GNE) at every time step \cite{WWTGS21,HBLD22,HB25,BG26,HBDL26,DFB26}. Each GT-MPC problem is a finite-horizon dynamic game; see~\cite{FRPDT20,LSM20,PB25} for solution algorithms and multi-vehicle applications.
Modeling the agents' costs is often difficult: the agents may be unwilling to disclose their objectives and strategies for privacy reasons, and in many applications the plant is even operated without explicit objectives to minimize, e.g., by standard decentralized control loops or even manually. A natural goal is then to learn, from the observed actions only, a GT-MPC law that reproduces them, so that the agents do not have to reveal their strategies, and that adds the main benefit of MPC, namely, handling input and output constraints. Moreover, once a game representation of agents' behavior is available, it can be used for higher-level tasks, such as optimization
of the set-point signals that enter the game at each step. It also allows reproducing the agents' behavior in closed-loop simulation, e.g., on a digital twin of the plant, to evaluate what-if scenarios that have not been observed yet.

In this paper we reach this goal by imitation: we learn a game whose equilibria, given reference signals and estimated states, reproduce the agents' actions.
We restrict our attention to strongly monotone linear-quadratic (LQ) games. Roughly speaking, monotonicity plays for games the role that convexity plays for optimization: if the joint feasible set is nonempty, closed, and convex, a strongly monotone game has a unique variational GNE for every value of the parameter (here, the state estimate, reference, and previous input)~\cite{FP03}. So the learned control law is well defined and, as LQ-GNE problems can be solved efficiently~\cite{ABB26,BT26b,Bem25nashopt}, easily computable. Moreover, the variational GNE of a strongly monotone LQ game is a path-differentiable function of the parameters, whose derivatives can be computed by automatic differentiation~\cite{BPS24}, which makes gradient-based learning and equilibrium design well founded.

Unlike~\cite{BGP15} and~\cite{AGD22}, who fit the agents' objectives by minimizing the residuals of the variational inequality and of the KKT conditions at observed equilibria, respectively, we impose strong monotonicity of the learned game by construction, as also done in~\cite{MBMG26} through an LMI. Moreover, in~\cite{BGP15,AGD22,MBMG26} the number of unknowns grows linearly with the number of samples, as the dual variables of each sample are optimized.
Finally, in our setting only the first move of each equilibrium is observed, not the entire sequence of planned moves that the methods in~\cite{BGP15,AGD22,MBMG26} would require to fit a GT-MPC formulation.
Learning was also combined with GT-MPC in~\cite{KZSB25} for two autonomous vehicles: the GNEs of a known constrained dynamic game are computed offline from random initial conditions, and a neural network trained to predict their reward outcome is used as the terminal cost of each agent's MPC problem. Here, instead, we learn the unknown costs from closed-loop data, do not assume decoupled dynamics, and solve the learned LQ game online.

We define a class of GT-MPC problems with soft shared output constraints based on a linear
state-space model realization learned from the available input / output data.
Then, we learn agent-specific weights and slack penalties by a bilevel problem whose inner problems are the LQ-GNE problems of the game at each datapoint, in which strong monotonicity holds by construction
and closed-loop stability of the unconstrained Nash equilibrium feedback is enforced by a penalty involving a jointly learned quadratic Lyapunov function, so that the problem can be solved by L-BFGS-B.
We illustrate in a numerical example that we can reconstruct an underlying game played by the agents
very precisely and shape closed-loop behavior by reference optimization.

Section~\ref{sec:setting} presents the problem setting,
Section~\ref{sec:lq-games} recalls monotone LQ games, Section~\ref{sec:pseudogradient} characterizes the monotonicity of the GT-MPC game, Section~\ref{sec:learning} presents the bilevel learning problem and its solution algorithm, Section~\ref{sec:stability} closed-loop stability,
Section~\ref{sec:example} a numerical example, and Section~\ref{sec:conclusions} conclusions.
The code is available at \url{https://github.com/bemporad/gtmpc_learn}.

\paragraph{Notation.} We write $I_n$ for the $n\times n$ identity, $\otimes$ for the Kronecker product, and $\sym(M)=\tfrac12(M+M\T)$ for the symmetric part of a square matrix $M$. For symmetric matrices, $M\succeq N$ means that $M-N$ is positive semidefinite. We write $\|\cdot\|_2$ for the Euclidean norm and $\|\cdot\|_F$ for the Frobenius norm. For a vector $x$ whose entries are associated with the agents, $x_i$ collects the entries of agent $i$ and $x_{-i}$ those of all the other agents. For a matrix $A$ acting on such vectors, $A_{ij}$ is the block that maps $x_j$ into the entries of agent $i$. We use $k$ for sample steps and $t$ for prediction steps. For a vector $a\in\rr^n$, $[a]_+=\max\{a,0\}$. For a differentiable vector- or matrix-valued function $f(\theta)$, $\mathrm df=\sum_l\frac{\partial f}{\partial\theta_l}\mathrm d\theta_l$ denotes its differential. For a matrix $G$, a vector $b$, and a set of indices $\mathcal S$, $G_{\mathcal S}$ is the submatrix of the rows of $G$ indexed by $\mathcal S$, and $b_{\mathcal S}$ the subvector of $b$ indexed by $\mathcal S$.

\section{Data collection and problem setup}
\label{sec:setting}

We consider a multi-agent process with $N$ agents, whose behavior we want to imitate, in which agent $i$ applies the input $u_{k,i}\in\rr^{n_i}$ at sample step $k$. We write $u_k=(u_{k,1},\ldots,u_{k,N})\in\rr^{n_u}$, with $n_u=\sum_{i=1}^Nn_i$, and $y_k\in\rr^{n_y}$ for the measured outputs.
We record piecewise-constant references $r^j_k\in\rr^{n_y}$, inputs $u^j_k$, and outputs $y^j_k$ in $n_\mathrm{e}$ experiments, each of length $N_j$, where the input and output measurements can be noisy.

Our goal is to learn a GT-MPC controller from the collected data, as described in the sequel, which imitates the behavior of the agents in tracking their respective references while respecting the following additional input and output constraints:
\begin{equation}
u_k\in\mathcal U
\qquad
\Delta u_k\in\Delta\mathcal U
\qquad
y_k\in\mathcal Y
\label{eq:io-constraints}
\end{equation}
where $\Delta u_k=u_k-u_{k-1}$ is the vector of input increments and
$\mathcal U\subseteq\rr^{n_u}$, $\Delta\mathcal U\subseteq\rr^{n_u}$, $\mathcal Y\subseteq\rr^{n_y}$ are polyhedral sets. For simplicity, we assume that $\mathcal U=\{u\mid u_{\min}\leq u\leq u_{\max}\}$,
$\Delta\mathcal U=\{\Delta u\mid \Delta u_{\min}\leq \Delta u\leq \Delta u_{\max}\}$,
$\mathcal Y=\{y\mid y_{\min}\leq y\leq y_{\max}\}$, with possibly infinite entries in $u_{\min}$, $u_{\max}$, $y_{\min}$, and $y_{\max}$. We denote by $u_{i,\min}$, $u_{i,\max}$, $\Delta u_{i,\min}$, and $\Delta u_{i,\max}$ the entries of $u_{\min}$, $u_{\max}$, $\Delta u_{\min}$, and $\Delta u_{\max}$ associated with agent $i$. As in standard MPC practice, constraints on the inputs and their increments are enforced as hard constraints, while output constraints are treated as soft constraints.

Our aim is to learn a GT-MPC controller of the following form, generalizing the one suggested in~\cite[Section~5.2]{Bem25nashopt}:
\begin{equation}
\begin{array}{ll}
\mbox{min}_{x_i,\epsilon_i} & \hspace*{-1em}\displaystyle\sum_{t=0}^{T-1}\bigl(y(t+1)-r_k\bigr)\T Q_{y,i}\bigl(y(t+1)-r_k\bigr)+\sum_{t=0}^{T_u-1}\Delta u(t)\T Q_{\Delta u,i}\Delta u(t)\\
&+q_{\epsilon,i}\epsilon_i+Q_{\epsilon,i}\epsilon_i^2\\[2pt]
\mbox{subject to} & \xi(t+1)={\mathcal A}\xi(t)+\sum_{i=1}^N{\mathcal B}_i u_i(t),\quad t=0,\ldots,T-1\\&\xi(0)=\hat\xi_k\\[2pt]
&y(t)={\mathcal C}\xi(t),\quad t=1,\ldots,T\\[2pt]
& u_{i,\min}\leq u_i(t)\leq u_{i,\max},\quad t=0,\ldots,T_u-1\\
&\Delta u_{i,\min}\leq\Delta u_i(t)\leq\Delta u_{i,\max},\quad t=0,\ldots,T_u-1\\[2pt]
& y_{\min}-\bigl(\textstyle\sum_{j=1}^N\epsilon_j\bigr)\mathbf 1\leq y(t)\leq y_{\max}+\bigl(\textstyle\sum_{j=1}^N\epsilon_j\bigr)\mathbf 1,\quad t=1,\ldots,T_c\\[2pt]
& \epsilon_i\geq0,\quad i=1,\ldots,N
\end{array}
\label{eq:gtmpc}
\end{equation}
where $\hat \xi_k$ is the estimate at sample step $k$ of the state vector $\xi\in\rr^n$ associated with the prediction model $(\mathcal A,\mathcal B,\mathcal C)$, with $\mathcal B=[\mathcal B_1\ \cdots\ \mathcal B_N]$ partitioned by agent according to the input allocation, $\Delta u(t)=u(t)-u(t-1)$ with $u(-1)=u_{k-1}$, and $x_i=\{\Delta u_i(t)\}_{t=0}^{T_u-1}$, $1\leq T_u\leq T$, with $\Delta u(t)=0$ for $t=T_u,\ldots,T-1$,
represents the sequence of input increments decided by agent $i$, $\epsilon_i$ is a slack variable for softening the shared output constraints, and $T$, $T_u\leq T$, and $T_c\leq T$ are, respectively, the prediction, control, and constraint horizons. Note that, in general, the agent-specific weights $Q_{y,i}$ and $Q_{\Delta u,i}$ consider the entire vectors of tracking-errors and input-increments, respectively.

Problem~\eqref{eq:gtmpc} can be recast as a parametric LQ-GNE problem
with parameter
$p_k$ $=$ $\bigl(\hat\xi_k,u_{k-1}$, $r_k\bigr)\in\rr^{n_p}$,
$n_p=n+n_u+n_y$,
and decision vector $z=(x,\epsilon)$, where $x=(x_1,\ldots,x_N)\in\rr^{n_uT_u}$ collects the input increments of all agents and $\epsilon=(\epsilon_1,\ldots,\epsilon_N)$. By letting $z^\star(p_k)=(x^\star(p_k),\epsilon^\star(p_k))$ denote the variational GNE of~\eqref{eq:gtmpc} corresponding to $p_k$, the MPC action at sample step $k$ is given by the first input increment of the optimal sequence, i.e.,
$u_{k}^{\textrm{MPC}}=u_{k-1}+E_\mathrm fx^\star(p_k)$, where $E_\mathrm f\in\{0,1\}^{n_u\times n_uT_u}$ selects the first increments $\Delta u(0)=(\Delta u_1(0),\ldots,\Delta u_N(0))$ from $x$.

As we only have input/output data available, we first need to identify a linear discrete-time model $(\mathcal A,\mathcal B,\mathcal C)$, together with the initial state $\hat\xi^j_0$ of each experiment, by solving the standard output-error minimization problem
\[
\begin{array}{ll}
\mbox{min}_{\mathcal A, \mathcal B, \mathcal C, \hat\xi^1_0,\ldots,\hat\xi^{n_\mathrm{e}}_0} & \displaystyle\sum_{j=1}^{n_\mathrm{e}}\sum_{k=0}^{N_j-1}\|y^j_k-\mathcal C\xi^j_k\|_2^2+\phi(\mathcal A,\mathcal B,\mathcal C,\hat\xi^1_0,\ldots,\hat\xi^{n_\mathrm{e}}_0)\\[2pt]
\mbox{subject to} & \xi^j_{k+1}=\mathcal A\xi^j_k+\mathcal Bu^j_k,\quad \xi^j_0=\hat\xi^j_0,\quad j=1,\ldots,n_\mathrm{e}
\end{array}
\]
where $\phi$ is a small $\ell_2$ regularization function, as described in~\cite{Bem25} and the associated \texttt{jax-sysid} package. Then, we run a Kalman filter based on the resulting model, whose noise covariances are tuning parameters, to get state estimates $\{\hat\xi^j_k\}_{k=0}^{N_j-1}$, starting from $\hat\xi^j_0$, given the corresponding inputs and outputs in experiment $j$.

The obtained datasets $\{(r^j_k,u^j_k,\hat\xi^j_k)\}$ are used for learning a parametric  linear-quadratic GNE problem of the form~\eqref{eq:gtmpc} by fitting the output and input-increment weights $Q_{y,i}$, $Q_{\Delta u,i}$, and the weights on slack variables $q_{\epsilon,i}$, $Q_{\epsilon,i}$ to best replicate the observed behavior of the expert controller, i.e., we wish that
\begin{equation}
    u^j_{k-1}+E_\mathrm f\,x^\star(p^j_k)\approx u^j_k
\label{eq:imitation}
\end{equation}
where $x^\star(p^j_k)$ is the vector of input increments of the variational GNE of~\eqref{eq:gtmpc} for $p=p^j_k$, i.e., we interpret the measured moves $u^j_k$ of the agents as the first moves of the GNE of a game of the form~\eqref{eq:gtmpc} played by the agents at step $k$.
Once the game is learned, the GT-MPC law applies the first move of the GNE, at which all the agents play best responses to each other. We also require that the resulting game be strongly monotone, so that its variational GNE, and hence the left-hand side of~\eqref{eq:imitation}, is unique whenever~\eqref{eq:gtmpc} is feasible\footnote{Infeasibility can only occur if $u_{k-1}$ and the imposed ranges ${\mathcal U},\Delta {\mathcal U}$ are incompatible.}.

\section{Learning monotone GT-MPC games}
\label{sec:lq-games}
In a parametric LQ game with constant agents' Hessian matrices and affine dependence
on the parameter $p$ in the linear cost terms, agent $i$ minimizes over $z_i=(x_i,\epsilon_i)$ the cost
\begin{equation}
J_i(z;p)=\tfrac12z_i\T\bar A_{ii}z_i+\sum_{j\neq i} z_i\T\bar A_{ij} z_j+\bar q_i(p)\T z_i+(\text{terms not depending on }z_i)
\label{eq:lq-cost}
\end{equation}
where $z=(x,\epsilon)$, $\bar A_{ii}$ is symmetric, and $\bar q_i(p)$ is affine in $p$. Agent $i$'s problem is convex when $\bar A_{ii}\succeq0$. The \emph{pseudogradient} $F(z;p)$ collects the agents' gradients $\nabla_{z_i}J_i(z;p)$ with respect to their own decision variables, in the same ordering as $z$,
so that $F(z;p)=\bar Az+\bar q(p)$,
with $\bar A=[\bar A_{ij}]\in\rr^{n_z\times n_z}$ and $\bar q$ collecting the vectors $\bar q_i$, where $n_z=n_uT_u+N$.
The game is \emph{monotone with constant} $\mu\geq0$~\cite{FP03,SPFP10}
if $(F(z;p)-F(z';p))\T(z-z')\geq\mu\|z-z'\|_2^2$ for all $z,z'$, which here is the same as
\begin{equation}
    \sym(\bar A)\succeq\mu I_{n_z}.
\label{eq:sym_A_succ_0}
\end{equation}
When $\mu>0$ the game is \emph{strongly monotone}.
With a shared constraint $z\in\mathcal Z(p)$, where $\mathcal Z(p)$ is nonempty, closed, and convex, the variational GNEs are the solutions of the variational inequality defined by $F(\cdot;p)$ on $\mathcal Z(p)$~\cite{FK07}. Since $F$ is continuous and strongly monotone, this variational inequality has a unique solution~\cite{FP03}, so the game has a unique variational GNE.

\subsection{Pseudogradient structure}
\label{sec:pseudogradient}
Let $u_\mathrm{full}=(u(0),\ldots,u(T-1))\in\rr^{n_uT}$
denote the input sequence for the full horizon, which is related to the increments $x_i\in\rr^{n_iT_u}$ of the agents as
$u_\mathrm{full}=(\mathbf 1_T\otimes I_{n_u})\,u_{k-1}+\sum_{i=1}^N\Lambda_ix_i$, $\Lambda_i=\Pi\otimes\bar E_i\T$,
where $\mathbf 1_T\in\rr^T$ is the vector of ones, $\Pi\in\{0,1\}^{T\times T_u}$ has entries $\Pi_{hl}=1$ if $l\leq h$ and $0$ otherwise, and $\bar E_i\in\{0,1\}^{n_i\times n_u}$ selects $u_i=\bar E_iu$. Then, the explicit form of the predicted outputs
$Y=(y(1),\ldots,y(T))\in\rr^{n_yT}$ is the following affine function of $(p,x)$:
\[
Y=\Phi\,\hat\xi_k+\Gamma u_\mathrm{full}
\qquad
\Phi=\begin{bmatrix}\mathcal C\mathcal A\\\mathcal C\mathcal A^2\\\vdots\\\mathcal C\mathcal A^T\end{bmatrix}
\qquad
\Gamma=\begin{bmatrix}
\mathcal C\mathcal B&0&\cdots&0\\
\mathcal C\mathcal A\mathcal B&\mathcal C\mathcal B&\cdots&0\\
\vdots&&\ddots&\vdots\\
\mathcal C\mathcal A^{T-1}\mathcal B&\mathcal C\mathcal A^{T-2}\mathcal B&\cdots&\mathcal C\mathcal B
\end{bmatrix}
\]
with $\Phi\in\rr^{n_yT\times n}$ and $\Gamma\in\rr^{n_yT\times n_uT}$ or, equivalently,
$Y=\Phi\,\hat\xi_k+\Gamma_uu_{k-1}+\sum_{i=1}^N\Gamma_ix_i$, where
$\Gamma_i=\Gamma\Lambda_i\in\rr^{n_yT\times n_iT_u}$ and
$\Gamma_u=\Gamma(\mathbf 1_T\otimes I_{n_u})\in\rr^{n_yT\times n_u}$.
By substituting the predictions in~\eqref{eq:gtmpc}, the cost of agent $i$ is an LQ cost of the form~\eqref{eq:lq-cost}, with $\bar A_{ii}=\blkdiag(A_{ii},2Q_{\epsilon,i})$, $\bar A_{ij}=\blkdiag(A_{ij},0)$ for $j\neq i$, and $\bar q_i(p)=(q_i(p),q_{\epsilon,i})$, where
\begin{equation}
\begin{aligned}
A_{ij}&=2\,\Gamma_i\T\bar Q_{y,i}\Gamma_j+2\,I_{T_u}\otimes Q^{ij}_{\Delta u,i}\\
q_i(p_k)&=2\,\Gamma_i\T\bar Q_{y,i}\bigl(\Phi\,\hat\xi_k+\Gamma_uu_{k-1}-(\mathbf 1_T\otimes I_{n_y})\,r_k\bigr)
\end{aligned}
\label{eq:pseudogradient-matrix-mpc}
\end{equation}
for all $i,j$, with $\bar Q_{y,i}=I_T\otimes Q_{y,i}$, $Q^{ij}_{\Delta u,i}=\bar E_iQ_{\Delta u,i}\bar E_j\T\in\rr^{n_i\times n_j}$. So $q_i(p)$ is linear in $p$. Only the block row $i$ of $Q_{\Delta u,i}$ enters $A$, i.e., agent $i$'s own weight $Q^{ii}_{\Delta u,i}$ and the cross weights $S_i=\bar E_iQ_{\Delta u,i}\bar E_{-i}\T\in\rr^{n_i\times(n_u-n_i)}$, where $\bar E_{-i}$ selects $u_{-i}$; the other blocks of $Q_{\Delta u,i}$ multiply terms that do not depend on $x_i$, and do not affect the game.

The row block $i$ of $A$ and $q$ depends only on agent $i$'s weights $Q_{y,i}$ and $Q_{\Delta u,i}$.
For a fixed model $(\mathcal A,\mathcal B,\mathcal C)$, both $A$ and $q$ are linear in the weight matrices. The slack $\epsilon_i$ enters agent $i$'s problem only through the terms $q_{\epsilon,i}\epsilon_i+Q_{\epsilon,i}\epsilon_i^2$ and the shared output constraints.
Note that the diagonal block $A_{ii}=2\Gamma_i\T\bar Q_{y,i}\Gamma_i+2(I_{T_u}\otimes Q^{ii}_{\Delta u,i})$ is positive definite when $Q_{y,i}\succeq0$ and $Q^{ii}_{\Delta u,i}\succ0$. Agent $i$'s problem~\eqref{eq:gtmpc} is then a strictly convex QP in $x_i$, and also in $\epsilon_i$ when $Q_{\epsilon,i}>0$.

The pseudogradient of the GT-MPC game is
$F(z;p)=\bigl(Ax+q(p),\,q_\epsilon+2Q_\epsilon\epsilon\bigr)$, with $A=[A_{ij}]\in\rr^{n_uT_u\times n_uT_u}$, $q=(q_1,\ldots,q_N)$,
$q_\epsilon=(q_{\epsilon,1},\ldots,q_{\epsilon,N})$ and
$Q_\epsilon=\mathrm{diag}(Q_{\epsilon,1},\ldots,Q_{\epsilon,N})$,
whose Jacobian with respect to $z$ is $\bar A=\blkdiag(A,2Q_\epsilon)$, and $\bar q(p)=(q(p),q_\epsilon)$. So, by~\eqref{eq:sym_A_succ_0}, the game is
strongly monotone in $z$, with constant $\mu>0$, if and only if
\begin{equation}
\sym(A)\succeq\mu I_{n_uT_u}
\qquad
Q_{\epsilon,i}\geq\frac\mu2,\quad i=1,\ldots,N.
\label{eq:strong-monotonicity}
\end{equation}
For a fixed model, the first condition is an LMI in $(Q_{y,i},Q_{\Delta u,i})_{i=1}^N$, and the second one is a simple bound on each quadratic slack penalty.

\begin{proposition}%
\label{prop:monotone-mpc}
Suppose that~\eqref{eq:strong-monotonicity} holds with $\mu>0$, that $u_{k-1}\in\mathcal U$, and that the bounds on the increments contain $0$. Then, for every $p$, the GT-MPC game has a unique variational GNE $z^\star(p)=(x^\star(p),\epsilon^\star(p))$.
\end{proposition}
\begin{proof}
The shared feasible set is closed and convex, and it is nonempty, since $x=0$ with large enough slacks is feasible. By~\eqref{eq:strong-monotonicity}, $F(\cdot;p)$ is continuous and strongly monotone with constant $\mu$ on this set, so the variational inequality that it defines on the set has a unique solution~\cite{FP03}, which is the unique variational GNE.
\end{proof}

\subsection{Learning problem}
\label{sec:learning}

Our goal now is to set up a learning problem that identifies all the weights in~\eqref{eq:gtmpc}
such that the resulting game is strongly monotone, i.e., satisfies~\eqref{eq:strong-monotonicity}. Since the pseudogradient matrix of the GT-MPC game is constrained to the structure~\eqref{eq:pseudogradient-matrix-mpc}, we cannot directly exploit the form $A=W\T W+D-D\T+\mu I_{n_uT_u}$ that
characterizes every strongly monotone LQ game, as was recently shown in~\cite[Lemma 4.4]{BT26c},
where $W$ is upper triangular and $D$ is block strictly upper triangular.
Instead, let us parametrize the agent-specific weights with variables $C_{y,i}\in\rr^{n_y\times n_y}$, $C_i\in\rr^{n_i\times n_i}$, and $S_i\in\rr^{n_i\times(n_u-n_i)}$, $i=1,\ldots,N$, where $Q_{y,i}=C_{y,i}\T C_{y,i}$, $S_i$ are the cross weights on input increments, and the own weights are
\begin{equation}
Q^{ii}_{\Delta u,i}=C_i\T C_i+\bigl(\eta+c(\theta)\bigr)I_{n_i}
\qquad
c(\theta)=\tfrac12\bigl[\mu-\lambda_{\min}\bigl(\sym(\tilde A(\theta))\bigr)\bigr]_+
\label{eq:shift}
\end{equation}
where $\eta>0$ is small and fixed, while $\tilde A(\theta)$ is the pseudogradient matrix~\eqref{eq:pseudogradient-matrix-mpc} obtained with $c(\theta)=0$. By~\eqref{eq:pseudogradient-matrix-mpc}, adding $cI_{n_i}$ to each  $Q^{ii}_{\Delta u,i}$ provides $\lambda_{\min}(\sym(A(\theta)))\geq\mu$ for every $\theta$.
The remaining block $\bar E_{-i}Q_{\Delta u,i}\bar E_{-i}\T$ does not affect the game; after learning we set it to $S_i\T(Q^{ii}_{\Delta u,i})^{-1}S_i$, the smallest block for which $Q_{\Delta u,i}\succeq0$. The slack penalties $q_{\epsilon,i}$ and $Q_{\epsilon,i}$ are scalars and are kept directly as optimization variables. Then, we learn the agent-specific weights and the slack penalties by solving
\begin{equation}
\begin{array}{ll}
\mbox{min}_\theta & \displaystyle\frac1{NK}\sum_{k=1}^K\bigl\|
\bar u_k(\theta)-u_k\bigr\|_2^2+\frac\rho2\sum_{i=1}^N\bigl(\|Q_{y,i}\|_F^2+\|Q^{ii}_{\Delta u,i}\|_F^2+2\|S_i\|_F^2\bigr)\\[2pt]
\mbox{subject to} & q_{\epsilon,i}\geq0,\quad Q_{\epsilon,i}\geq\dfrac\mu2,\quad i=1,\ldots,N
\end{array}
\label{eq:learning_problem}
\end{equation}
where $\theta=(C_{y,i},C_i,S_i,q_{\epsilon,i},Q_{\epsilon,i})_{i=1}^N$ is the vector that collects all the learnable parameters of the game, $u_k$ is the measured input at sample step $k$, and $\bar u_k(\theta)=u_{k-1}+E_\mathrm f\,x^\star(p_k;\theta)$ is the first move of the variational GNE 
\[
    z^\star(p_k;\theta)=(x^\star(p_k;\theta),\epsilon^\star(p_k;\theta)) 
\]
of the game~\eqref{eq:gtmpc} with the weights $\theta$ at the parameter $p_k$. The scalar $\rho>0$ weights the regularization of the weights $Q_{y,i}$, $Q^{ii}_{\Delta u,i}$, and $S_i$.

Due to the GNE terms $\bar u_k(\theta)$, the optimization problem in~\eqref{eq:learning_problem} is a bilevel problem, whose lower-level problems are the GNE problems of the game at the parameters $p_k$. By Proposition~\ref{prop:monotone-mpc}, $\bar u_k(\theta)$ is uniquely defined for all $k$ and admissible $\theta$, as the measured inputs satisfy the input constraints. For compactness of notation,
in~\eqref{eq:learning_problem} we drop the superscript $j$ and number all the sample steps of all experiments consecutively as $k=1,\ldots,K$. We describe below how we solve this bound-constrained bilevel nonlinear least-squares problem with L-BFGS-B~\cite{BLNZ95}, including the initial guess and the computation of the GNEs and of their derivatives with respect to $\theta$.

\paragraph{Initialization.}
We initialize the components of $\theta$ as
\[
\begin{aligned}
C_{y,i}=&\sqrt s\,I_{n_y}
\quad
C_i=\sqrt{\tfrac{\mu+\delta}2-\eta}\,I_{n_i}
\quad
S_i=0
\quad 
q_{\epsilon,i}=q^0_\epsilon
\quad
Q_{\epsilon,i}=Q^0_\epsilon\\
&\quad
i=1,\ldots,N
\end{aligned}
\]
where $s\geq0$ is a scale factor, $\delta>0$ is small, $\eta\leq(\mu+\delta)/2$, $q^0_\epsilon>0$, and $Q^0_\epsilon\geq\mu/2$.
Then \eqref{eq:pseudogradient-matrix-mpc} gives the symmetric matrix $\tilde A(\theta)=2s\bar\Gamma\T\bar\Gamma+(\mu+\delta)I_{n_uT_u}\succeq(\mu+\delta)I_{n_uT_u}$ for every $s\geq0$, where $\bar\Gamma=[\Gamma_1\ \cdots\ \Gamma_N]$, so that $c(\theta)=0$ at the initial point. With the slack penalties fixed to $q^0_\epsilon$ and $Q^0_\epsilon$, we choose $s$ by minimizing the imitation loss, i.e., the first term of the objective in~\eqref{eq:learning_problem}, over a logarithmic grid of values of $s$, which requires solving $K$ LQ-GNE problems for each grid value.

\paragraph{Computing the GNEs and their derivatives.}
As L-BFGS-B is a gradient-based method, we need to differentiate the LQ-GNEs $z^\star(p_k;\theta)$ with respect to $\theta$ efficiently.
Recall that the pseudogradient of the game is $\bar A(\theta)z+\bar q(p)$, with $\bar A(\theta)=\blkdiag(A(\theta),2Q_\epsilon)$ and $\bar q(p)=(q(p),q_\epsilon)$, and let $Gz\leq b(p_k)$ collect the shared constraints of~\eqref{eq:gtmpc}, where $b$ also depends on $u_{k-1}$. As by strong monotonicity~\eqref{eq:sym_A_succ_0} matrix $\bar A(\theta)$ is invertible, by~\cite[Lemma~5.1]{BT26b} the variational GNE is $z^\star_k=z_{0,k}-\bar A^{-1}G\T\lambda^\star_k$, $z_{0,k}=-\bar A^{-1}\bar q(p_k)$, where $\lambda^\star_k$ solves the QP
\begin{equation}
\mbox{min}_{\lambda}\ \lambda\T\sym(H)\lambda+q_{0,k}\T\lambda\quad
\mbox{s.t.}\ H\lambda+q_{0,k}\geq0,\ \lambda\geq0
\label{eq:reduced-qp}
\end{equation}
with $H=G\bar A^{-1}G\T$ and $q_{0,k}=b(p_k)-Gz_{0,k}$. Although $\bar u_k(\theta)$ and its derivatives could be obtained directly by solving~\eqref{eq:reduced-qp} with a differentiable QP solver~\cite{AK17,TM24}, we take a more efficient approach here. Let $\theta^j$ be the value of the parameters at the $j$-th iteration of the learning algorithm. We compute $z^\star_k(\theta^j)$ by a tailored LQ-GNE solver (such as DR-DAQP~\cite{ABB26}) and identify the set $\mathcal A_k(\theta^j)$ of constraints with positive multipliers (we drop the dependence on $\theta^j$ from now on for simplicity of notation). Then, we write $z^\star_k=z_{0,k}-\bar A^{-1}G_{\mathcal A_k}\T\lambda_{\mathcal A_k}$, where $G_{\mathcal A_k}\bar A^{-1}G_{\mathcal A_k}\T\lambda_{\mathcal A_k}=G_{\mathcal A_k}z_{0,k}-b_{\mathcal A_k}(p_k)$ are the rows of $H\lambda+q_{0,k}=0$ of the active constraints. For a fixed active set, this is a linear system $H_{\mathcal A_k}(\theta)\lambda_{\mathcal A_k}=h_k(\theta)$, with $H_{\mathcal A_k}=G_{\mathcal A_k}\bar A^{-1}G_{\mathcal A_k}\T$ and $h_k=G_{\mathcal A_k}z_{0,k}-b_{\mathcal A_k}(p_k)$, whose coefficients depend smoothly on $\theta$ through $\bar A(\theta)$ and $\bar q(p_k)$. If the rows of $G_{\mathcal A_k}$ are linearly independent, the multipliers $\lambda_{\mathcal A_k}$ are positive, and the inactive constraints are strictly satisfied, the active set does not change for small perturbations of $\theta$. Then $\lambda_{\mathcal A_k}(\theta)$, $z^\star_k(\theta)$, and hence $\bar u_k(\theta)$ are smooth functions of $\theta$ around $\theta^j$, with $\mathrm d\lambda_{\mathcal A_k}=H_{\mathcal A_k}^{-1}(\mathrm d h_k-\mathrm dH_{\mathcal A_k}\lambda_{\mathcal A_k})$, obtained by differentiating $H_{\mathcal A_k}\lambda_{\mathcal A_k}=h_k$, and hence $\mathrm dz^\star_k=\mathrm dz_{0,k}-\mathrm d(\bar A^{-1})\,G_{\mathcal A_k}\T\lambda_{\mathcal A_k}-\bar A^{-1}G_{\mathcal A_k}\T\mathrm d\lambda_{\mathcal A_k}$.
In our numerical experiments, we implement the map from $\theta$ to the imitation loss through these linear systems in \texttt{JAX}~\cite{JAX}.

\subsection{Closed-loop stability enforcement}
\label{sec:stability}
Strong monotonicity guarantees that the GT-MPC law is well defined, but not that the closed loop is asymptotically stable. Without constraints, the variational GNE is $x^\star(p)=-A(\theta)^{-1}q(p)$, which is linear in $p=(\hat\xi_k,u_{k-1},r_k)$, so that the vector of first input increments of all agents is $K_\xi(\theta)\hat\xi_k+K_u(\theta)u_{k-1}+K_r(\theta)r_k$.
In closed loop with the identified model and $\hat\xi_k=\xi_k$, the state $\zeta_k=(\xi_k,u_{k-1})$ evolves as $\zeta_{k+1}=M(\theta)\zeta_k+M_r(\theta)r_k$, for some matrix $M_r(\theta)$, where
$
M(\theta)=\left[\begin{smallmatrix}{\mathcal A}+{\mathcal B}K_\xi(\theta) & {\mathcal B}\bigl(I+K_u(\theta)\bigr)\\ K_\xi(\theta) & I+K_u(\theta)\end{smallmatrix}\right]$.
Under the assumption that the observed multi-agent system generating the training data is asymptotically stable,
we can enforce asymptotic stability of the unconstrained closed loop a priori by adding to the objective of~\eqref{eq:learning_problem}
the following penalty
\begin{equation}
\rho_s\bigl[\|LM(\theta)L^{-1}\|_2^2-\gamma^2\bigr]_+^2
\label{eq:stability-penalty}
\end{equation}
where $\rho_s>0$ weights the penalty relative to the imitation loss,
$\gamma\in(0,1)$ is a desired contraction rate, and the matrix $L\in\rr^{(n+n_u)\times(n+n_u)}$, upper triangular with positive diagonal, is learned together with $\theta$. Since $\|LML^{-1}\|_2<1$ is equivalent to $M\T PM\prec P$ with $P=L\T L$, a value of $\|LM(\theta)L^{-1}\|_2$ below $1$ certifies the quadratic Lyapunov function $\zeta\T P\zeta$ of the unconstrained closed loop. Conversely, by~\cite[Lemma~1]{Bem25} applied to $M/\gamma$, for every matrix $M$ with spectral radius smaller than $\gamma$ there exists such an $L$, given by the Cholesky factor of the solution of a Lyapunov equation, for which $\|LML^{-1}\|_2<\gamma$, so the penalty can be zeroed for every game whose closed-loop matrix $M(\theta)$ has spectral radius smaller than $\gamma$. As~\eqref{eq:stability-penalty} is a penalty, we check that $\|LM(\theta)L^{-1}\|_2<1$ at the solution and, otherwise, increase $\rho_s$ and solve the problem again starting from the current solution. By the separation principle, stability of $M(\theta)$ also implies stability of the nominal closed loop with the Kalman filter when constraints are not active, although it does not guarantee global closed-loop stability under constraints, nor robustness to the mismatch between the identified model and the actual process.

\section{Numerical example}
\label{sec:example}

\paragraph{Generating system.} We consider a randomly generated open-loop stable process with $n_u=3$ inputs, $n_y=3$ outputs, $n=4$ states, and $N=3$ agents, each commanding one input. The constraints are $|u_i|\leq0.15$, $|\Delta u_i|\leq0.1$, and we want to enforce $|y_i|\leq y_{\textrm{max}}=3$. The agents operate the process by playing a GT-MPC game of the form~\eqref{eq:gtmpc} with $T=T_c=10$, $T_u=4$, the true model and state, the soft output constraints $|y_i|\leq3$ with the slack penalties $q_{\epsilon,i}=0.003$ and $Q_{\epsilon,i}=0.5$, and the weights
$
Q_{y,1}=10^{-4}\mathrm{diag}(1,0.6,0.1)$,
$
Q_{y,2}=10^{-4}\mathrm{diag}(0.1,1,0.6)$,
$
Q_{y,3}=10^{-4}\mathrm{diag}(0.6,0.1,1)
$,
and $Q_{\Delta u,i}=w_ie_ie_i\T$, where $e_i$ is the $i$th column of $I_3$ and $(w_1,w_2,w_3)=(0.05,0.1,0.2)$.
The pseudogradient matrix $A$ of this game is not symmetric, $\|A-A\T\|_F/\|A\|_F=0.30$, so the game is not a potential game that can be represented by a single centralized MPC cost. It is also strongly monotone, with $\lambda_{\min}(\sym(A))=0.106$. All three inputs and outputs are dynamically coupled by the system dynamics.

\paragraph{Data collection.} We record $n_\mathrm{e}=7$ experiments of $200$ steps in which the agents operate according to the underlying (unknown) GT-MPC strategy. Each experiment starts from the equilibrium of a reference uniformly distributed in $[-1.5,1.5]^3$, a range in which the corresponding equilibrium input always satisfies the input constraints, and the references of all the agents are piecewise constant, with new values uniformly distributed in $[-5,5]^3$ every $15$ steps. The output noise is uniformly distributed in $[-0.05,0.05]$.

\paragraph{Learning.} Using \texttt{jax-sysid}, we first identify a model with $n=4$ states from these closed-loop data in $2.3$~s\footnote{All the reported CPU times were obtained on an Apple M5 Max machine in Python 3.11.15.} . On $3$ further experiments generated in the same way, the open-loop simulation of the identified model, from the initial states estimated by \texttt{jax-sysid}, achieves best fit rates $100\bigl(1-\|y_l-\hat y_l\|_2/\|y_l-\bar y_l\|_2\bigr)$ of $98.7\%$, $98.6\%$, and $98.7\%$ on the outputs $l=1,2,3$, where $y_l$ collects the measured values of output $l$, $\hat y_l$ the simulated ones, and $\bar y_l$ is the mean of $y_l$.

For the GT-MPC structure to learn, we use $T=T_c=10$, $\mu=0.1$, $\rho=10^{-5}$, $\eta=10^{-3}$, and
different values of $T_u$ between 1 and $T$. We initialize the bilevel problem with $q^0_\epsilon=10^{-3}$, $Q^0_\epsilon=0.1$, and $\delta=0.01$, searching $s$ over the grid $\log_{10}s\in\{-10,-9.5,\ldots,-2\}$
for initializing the solution of~\eqref{eq:learning_problem}, which we solve by the L-BFGS-B solver of \texttt{SciPy} with the stability penalty~\eqref{eq:stability-penalty}, $\gamma=0.9$, and $\rho_s=10$.
The Kalman filter uses the process noise covariance $10^{-6}I_4$, the measurement noise covariance $\frac{0.05^2}{3}I_3$,
and the initial covariance $10^{-4}I_4$.

\paragraph{Controller.} At each time step $k$, the GT-MPC law estimates $\hat\xi_k$ by a Kalman filter based on the identified model, sets $p_k=(\hat\xi_k,u_{k-1},r_k)$, and computes the variational GNE of~\eqref{eq:gtmpc}, which exists and is unique by Proposition~\ref{prop:monotone-mpc}, by the DR-DAQP solver~\cite{ABB26} through the \texttt{nashopt} package~\cite{Bem25nashopt}.

\paragraph{Results.} We validate the controllers on $10$ runs of $80$ steps, each with a set-point change from $r^0$ to $r^1$, starting from the equilibrium of $r^0$. Each component of $r^1$ is uniformly distributed in $[-2.7,2.7]$, and $r^0=r^1-\mathrm{sign}(r^1)\circ d$, with $d$ uniformly distributed in $[0,3]^3$.
The Kalman filter of the GT-MPC law starts from the equilibrium state of the identified model for $r^0$.
Table~\ref{tab:Tu} reports, for different values of $T_u$, the training time, the monotonicity constant $\lambda_{\min}(\sym(A(\theta)))$ of the learned game,
and the closed-loop imitation errors
\[
\begin{aligned}
\textrm{RMS}_u = &\biggl(\frac{\sum_{j=1}^{10}\sum_{k=0}^{79}\|u^j_k-u^j_{\mathrm{a},k}\|_2^2}{\sum_{j=1}^{10}\sum_{k=0}^{79}\|u^j_{\mathrm{a},k}\|_2^2}\biggr)^{1/2}\\
\textrm{RMS}_y = &\biggl(\frac{\sum_{j=1}^{10}\sum_{k=0}^{79}\|y^j_k-y^j_{\mathrm{a},k}\|_2^2}{\sum_{j=1}^{10}\sum_{k=0}^{79}\|y^j_{\mathrm{a},k}\|_2^2}\biggr)^{1/2}
\end{aligned}
\]
where $(y^j_k,u^j_k)$ and $(y^j_{\mathrm{a},k},u^j_{\mathrm{a},k})$ are the measured outputs and the inputs of the closed loops with the GT-MPC law and with the agents in run $j$.
The average CPU time per step taken by the solver to evaluate the GT-MPC action ranges between $21$ and $96$~$\mu$s depending on the control horizon $T_u$.
With closed-loop stability enforced, the GT-MPC law learned by~\eqref{eq:learning_problem} imitates the agents with imitation errors below $0.071$ for all $T_u$, and the smallest for $T_u=4$, which is the control horizon indeed used by the agents generating the training data.

\begin{table}[ht]
\centering
\scriptsize
\setlength{\tabcolsep}{3pt}
\begin{tabular}{lll}
\toprule
& Agents & Learned, $T_u=4$\\
\midrule
$\mathrm{diag}(Q_{y,1})$ & $10^{-4}(1,0.6,0.1)$ & $0.89\cdot10^{-4}(1.04,0.58,0.08)$\\
$\mathrm{diag}(Q_{y,2})$ & $10^{-4}(0.1,1,0.6)$ & $0.89\cdot10^{-4}(0.12,1.03,0.61)$\\
$\mathrm{diag}(Q_{y,3})$ & $10^{-4}(0.6,0.1,1)$ & $0.89\cdot10^{-4}(0.57,0.09,1.01)$\\
$Q^{11}_{\Delta u,1}$ & $0.05$ & $0.89\cdot0.053$\\
$Q^{22}_{\Delta u,2}$ & $0.1$ & $0.89\cdot0.101$\\
$Q^{33}_{\Delta u,3}$ & $0.2$ & $0.89\cdot0.201$\\
$q_{\epsilon,i}$ & $0.003$ & $0.89\cdot0.0028$\\
$Q_{\epsilon,i}$ & $0.5$ & $0.89\cdot0.54$\\
\bottomrule
\end{tabular}
\hfill
\begin{tabular}{ccccc}
\toprule
$T_u$ & Time [s] & $\lambda_{\min}$ & RMS$_u$ & RMS$_y$\\
\midrule
$1$ & $7.8$ & $0.1908$ & $0.0453$ & $0.0706$\\
$2$ & $10.9$ & $0.1000$ & $0.0214$ & $0.0245$\\
$3$ & $14.5$ & $0.1000$ & $0.0106$ & $0.0133$\\
$4$ & $21.7$ & $0.1000$ & $0.0016$ & $0.0026$\\
$5$ & $19.6$ & $0.1000$ & $0.0051$ & $0.0050$\\
$6$ & $11.1$ & $0.1000$ & $0.0080$ & $0.0064$\\
$7$ & $25.8$ & $0.1000$ & $0.0065$ & $0.0065$\\
$8$ & $36.4$ & $0.1000$ & $0.0062$ & $0.0069$\\
$9$ & $35.1$ & $0.1000$ & $0.0058$ & $0.0070$\\
$10$ & $57.8$ & $0.1000$ & $0.0058$ & $0.0070$\\
\bottomrule
\end{tabular}
\caption{Numerical example. Left: true weights used by the agents and weights learned with $T_u=4$. Right: learning time, monotonicity constant $\lambda_{\min}=\lambda_{\min}(\sym(A(\theta)))$, and imitation errors for $T_u=1,\ldots,T$.}
\label{tab:Tu}
\end{table}

\begin{figure}[t!]
\centering
\includegraphics[width=\textwidth]{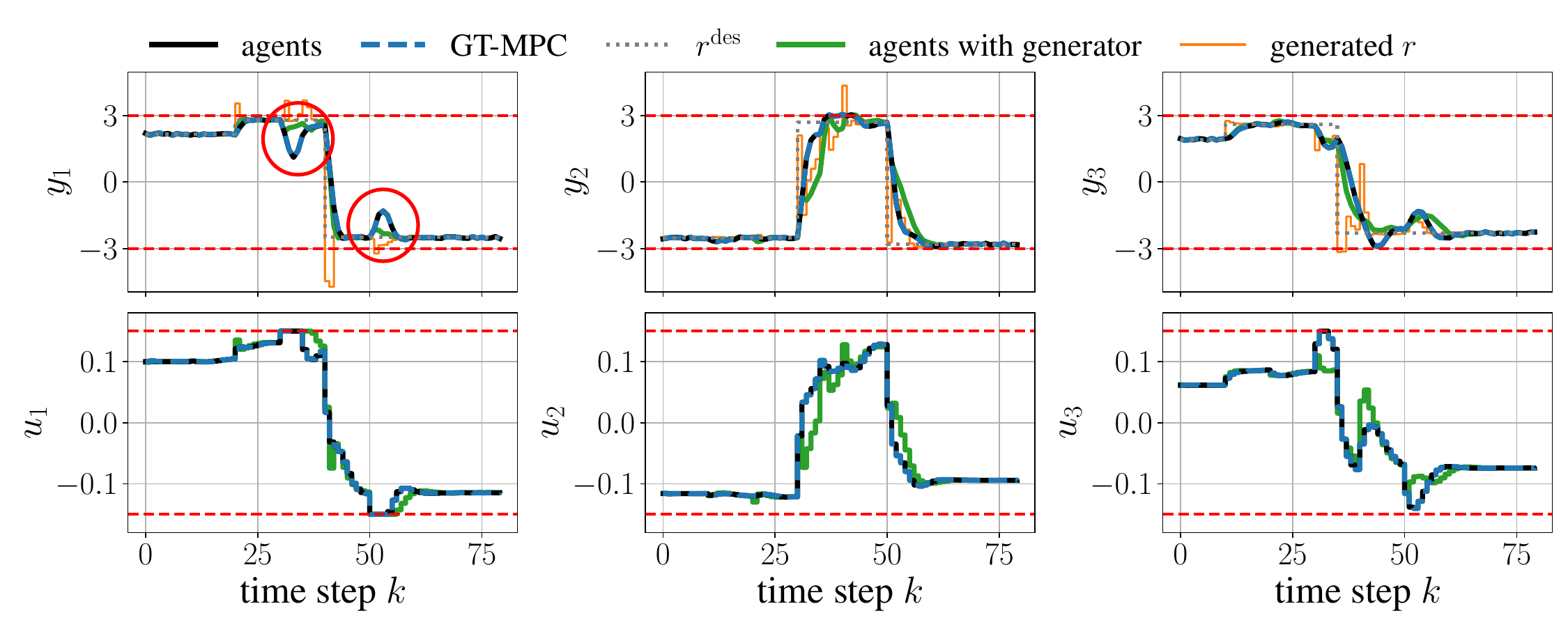}
\caption{Agents (black) and GT-MPC law learned with $T_u=4$ by~\eqref{eq:learning_problem} (blue, dashed), tracking the references $r^\mathrm{des}$ (dotted), agents reacting to the reference generator~\eqref{eq:ref-opt} (green), and generated references $r$ (orange). Top: outputs and bounds (dashed). Bottom: inputs.}
\label{fig:example}
\end{figure}

Figure~\ref{fig:example} shows a run of the GT-MPC law learned by~\eqref{eq:learning_problem} with $T_u=4$. The GT-MPC law produces closed-loop trajectories that are almost indistinguishable from those of the agents.
As shown in Table~\ref{tab:Tu} (left), for $T_u=4$ the learned weights are close to those of the agents up to a common scale factor $\alpha=0.89$,
which leaves the GNEs unchanged, and the off-diagonal entries of $Q_{y,i}$ and the cross weights $S_i$ are small, below $1.6\cdot10^{-6}$ and $2.8\cdot10^{-3}$ in magnitude, respectively.
With the stability penalty, the unconstrained closed loops are asymptotically stable for all $T_u$ (spectral radius between $0.72$ and $0.85$).

\subsection{Set-point optimization}
\label{sec:ref-opt}
The learned game can be used for higher-level tasks without requiring the agents to reveal their objectives. As an example, a
reference generator
placed on top of the agents, who keep playing their own GT-MPC game, sends them at each step $k$ the reference
\begin{equation}
r_k=\argmin_r\ \bigl(\hat y_1(r)-r^\mathrm{des}_{1,k}\bigr)^2+\omega\sum_{l=2}^3\bigl(\hat y_l(r)-r^\mathrm{des}_{l,k}\bigr)^2+\sigma\|r-r^\mathrm{des}_k\|_2^2
\label{eq:ref-opt}
\end{equation}
where $r^\mathrm{des}_k$ is the reference of Figure~\ref{fig:example} to track, $\hat y(r)=\mathcal C\bigl(\mathcal A\hat\xi_k+\mathcal B\bar u_k(r)\bigr)$ is the output predicted by the identified model when the agents apply the GT-MPC action $\bar u_k(r)$ of the game learned with $T_u=4$ at $p=(\hat\xi_k,u_{k-1},r)$, $\omega\in(0,1)$ reduces the importance of tracking $y_2$ and $y_3$, and $\sigma>0$ prevents excessive deviations of $r$ from $r^\mathrm{des}_k$. We solve~\eqref{eq:ref-opt} by a Gauss-Newton method, using the derivatives of the GNE of Section~\ref{sec:learning}. As shown in Figure~\ref{fig:example} (green) for $\omega=0.02$ and $\sigma=10^{-2}$, the generator largely removes the dip and the bump of $y_1$ caused by the coupling with the other loops (see red circles), at the price of slower transients of $y_2$ and $y_3$.

\section{Conclusions}
\label{sec:conclusions}
We have proposed a method for learning strongly monotone GT-MPC laws from input/output data of multiple agents, and illustrated the use of the learned game for explaining the agents' objectives and for optimizing the references given to the agents. Future research will address other higher-level uses of the learned game, such as mechanism design, and extensions to state-dependent weights and nonlinear systems.

\subsection*{Generative AI Use}

Generative AI tools were used to help write and proofread the paper and for generating the Python code used in the reported example.

\subsection*{Acknowledgments}
This work was funded by the European Union (ERC Advanced
Research Grant COMPACT, No. 101141351). Views and opinions expressed
are however those of the authors only and do not necessarily reflect
those of the European Union or the European Research Council. Neither
the European Union nor the granting authority can be held responsible
for them. The work was also supported by the Deutsche
Forschungsgemeinschaft (DFG, German Research Foundation), project
number 528033031.


\begin{thebibliography}{10}

\bibitem{AGD22}
S.~Allen, S.~A. Gabriel, and J.~P. Dickerson.
\newblock Using inverse optimization to learn cost functions in generalized {N}ash games.
\newblock {\em Computers \& Operations Research}, 142:105721, 2022.

\bibitem{AK17}
B.~Amos and J.~Z. Kolter.
\newblock {OptNet}: Differentiable optimization as a layer in neural networks.
\newblock In {\em Proceedings of the 34th International Conference on Machine Learning}, pages 136--145, 2017.

\bibitem{ABB26}
D.~Arnstr\"om, E.~Benenati, and G.~Belgioioso.
\newblock \texttt{DR-DAQP}: A hybrid operator splitting and active-set solver for affine variational inequalities.
\newblock arXiv preprint arXiv:2604.02531, 2026.

\bibitem{Bem25}
A.~Bemporad.
\newblock An {L-BFGS-B} approach for linear and nonlinear system identification under $\ell_1$ and group-lasso regularization.
\newblock {\em IEEE Transactions on Automatic Control}, 70(7):4857--4864, 2025.

\bibitem{Bem25nashopt}
A.~Bemporad.
\newblock {NashOpt}: A {P}ython library for computing generalized {N}ash equilibria.
\newblock arXiv preprint arXiv:2512.23636, 2025.

\bibitem{BT26c}
A.~Bemporad and T.~Tatarenko.
\newblock Learning parametric monotone games.
\newblock arXiv preprint arXiv:2609.02494, 2026.

\bibitem{BT26b}
A.~Bemporad and T.~Tatarenko.
\newblock Solving monotone linear-quadratic generalized {Nash} equilibrium problems via quadratic programming.
\newblock arXiv preprint arXiv:2608.07336, 2026.

\bibitem{BG26}
E.~Benenati and S.~Grammatico.
\newblock Linear-quadratic dynamic games as receding-horizon variational inequalities.
\newblock {\em IEEE Transactions on Automatic Control}, 71(4):2404--2417, 2026.

\bibitem{BGP15}
D.~Bertsimas, V.~Gupta, and I.C. Paschalidis.
\newblock Data-driven estimation in equilibrium using inverse optimization.
\newblock {\em Mathematical Programming}, 153(2):595--633, 2015.

\bibitem{BPS24}
J.~Bolte, E.~Pauwels, and A.~Silveti-Falls.
\newblock Differentiating nonsmooth solutions to parametric monotone inclusion problems.
\newblock {\em SIAM Journal on Optimization}, 34(1):71--97, 2024.

\bibitem{JAX}
J.~Bradbury, R.~Frostig, P.~Hawkins, M.J. Johnson, C.~Leary, D.~Maclaurin, G.~Necula, A.~Paszke, J.~Vander{P}las, S.~Wanderman-{M}ilne, and Q.~Zhang.
\newblock {JAX}: composable transformations of {P}ython+{N}um{P}y programs.
\newblock \url{http://github.com/google/jax}, 2018.

\bibitem{BLNZ95}
R.H. Byrd, P.~Lu, J.~Nocedal, and C.~Zhu.
\newblock A limited memory algorithm for bound constrained optimization.
\newblock {\em SIAM Journal on Scientific Computing}, 16(5):1190--1208, 1995.

\bibitem{DFB26}
E.~Dimou, F.~Fabiani, and A.~Bemporad.
\newblock Receding horizon games with stability guarantees.
\newblock In {\em 65th IEEE Conference on Decision and Control}, 2026.

\bibitem{FK07}
F.~Facchinei and C.~Kanzow.
\newblock Generalized {N}ash equilibrium problems.
\newblock {\em 4OR}, 5(3):173--210, 2007.

\bibitem{FP03}
F.~Facchinei and J.-S. Pang.
\newblock {\em Finite-Dimensional Variational Inequalities and Complementarity Problems}.
\newblock Springer, New York, 2003.

\bibitem{FRPDT20}
D.~Fridovich-Keil, E.~Ratner, L.~Peters, A.D. Dragan, and C.J. Tomlin.
\newblock Efficient iterative linear-quadratic approximations for nonlinear multi-player general-sum differential games.
\newblock In {\em Proc. IEEE International Conference on Robotics and Automation}, pages 1475--1481, Paris, France, 2020.

\bibitem{HBDL26}
S.~Hall, G.~Belgioioso, F.~D{\"o}rfler, and D.~Liao-McPherson.
\newblock Stability certificates for receding-horizon games.
\newblock {\em IEEE Transactions on Automatic Control}, 71(5):3511--3518, 2026.

\bibitem{HBLD22}
S.~Hall, G.~Belgioioso, D.~Liao-McPherson, and F.~D{\"o}rfler.
\newblock Receding horizon games with coupling constraints for demand-side management.
\newblock In {\em 61st IEEE Conference on Decision and Control}, pages 3795--3800, 2022.

\bibitem{HB25}
S.~Hall and A.~Bemporad.
\newblock Solving multiparametric generalized {Nash} equilibrium problems and explicit game-theoretic model predictive control.
\newblock arXiv preprint arXiv:2512.05505, 2025.
\newblock Code available at \url{https://github.com/bemporad/nash_mpqp}.

\bibitem{KZSB25}
H.~Kim, E.L. Zhu, C.S. Lim, and F.~Borrelli.
\newblock Learning two-agent motion planning strategies from generalized nash equilibrium for model predictive control.
\newblock In {\em Proc. 7th Annual Learning for Dynamics and Control Conference}, volume 283 of {\em Proceedings of Machine Learning Research}, pages 112--123, 2025.

\bibitem{LSM20}
S.~{Le Cleac'h}, M.~Schwager, and Z.~Manchester.
\newblock {ALGAMES}: A fast solver for constrained dynamic games.
\newblock In {\em Robotics: Science and Systems XVI}, 2020.

\bibitem{MBMG26}
A.~Mallick, R.~{Rahimi Baghbadorani}, P.~{Mohajerin Esfahani}, and S.~Grammatico.
\newblock Learning in inverse games: Tractable training with probabilistic guarantees.
\newblock {\em arXiv preprint arXiv:2610.03481}, 2026.

\bibitem{PB25}
M.~Pustilnik and F.~Borrelli.
\newblock Generalized {N}ash equilibrium solutions in dynamic games with shared constraints.
\newblock arXiv preprint arXiv:2502.19569, 2025.

\bibitem{SPFP10}
G.~Scutari, D.~P. Palomar, F.~Facchinei, and J.-S. Pang.
\newblock Convex optimization, game theory, and variational inequality theory.
\newblock {\em IEEE Signal Processing Magazine}, 27(3):35--49, 2010.

\bibitem{TM24}
K.~Tracy and Z.~Manchester.
\newblock On the differentiability of the primal-dual interior-point method.
\newblock arXiv preprint arXiv:2406.11749, 2024.

\bibitem{WWTGS21}
M.~Wang, Z.~Wang, J.~Talbot, J.~C. Gerdes, and M.~Schwager.
\newblock Game-theoretic planning for self-driving cars in multivehicle competitive scenarios.
\newblock {\em IEEE Transactions on Robotics}, 37(4):1313--1325, 2021.

\end{thebibliography}
\end{document}